\documentclass[twocolumn,english,aps,prb,floats]{revtex4}
\usepackage[T1]{fontenc}
\usepackage[latin9]{inputenc}
\usepackage{babel}
\usepackage{amsmath}
\usepackage{amssymb}
\usepackage{graphicx}
\usepackage{esint}
\usepackage[unicode=true,
 bookmarks=false,
 breaklinks=false,pdfborder={0 0 1},backref=section,colorlinks=false]
 {hyperref}

\makeatletter
\@ifundefined{textcolor}{}
{%
 \definecolor{BLACK}{gray}{0}
 \definecolor{WHITE}{gray}{1}
 \definecolor{RED}{rgb}{1,0,0}
 \definecolor{GREEN}{rgb}{0,1,0}
 \definecolor{BLUE}{rgb}{0,0,1}
 \definecolor{CYAN}{cmyk}{1,0,0,0}
 \definecolor{MAGENTA}{cmyk}{0,1,0,0}
 \definecolor{YELLOW}{cmyk}{0,0,1,0}
}

\usepackage{babel}
\usepackage{babel}

\usepackage{babel}
\usepackage{bm}
\usepackage{wasysym}
\usepackage{breakurl}

\@ifundefined{textcolor}{}{%
 \definecolor{BLACK}{gray}{0}
 \definecolor{WHITE}{gray}{1}
 \definecolor{RED}{rgb}{1,0,0}
 \definecolor{GREEN}{rgb}{0,1,0}
 \definecolor{BLUE}{rgb}{0,0,1}
 \definecolor{CYAN}{cmyk}{1,0,0,0}
 \definecolor{MAGENTA}{cmyk}{0,1,0,0}
 \definecolor{YELLOW}{cmyk}{0,0,1,0}
}

\@ifundefined{textcolor}{}{%
 \definecolor{BLACK}{gray}{0}
 \definecolor{WHITE}{gray}{1}
 \definecolor{RED}{rgb}{1,0,0}
 \definecolor{GREEN}{rgb}{0,1,0}
 \definecolor{BLUE}{rgb}{0,0,1}
 \definecolor{CYAN}{cmyk}{1,0,0,0}
 \definecolor{MAGENTA}{cmyk}{0,1,0,0}
 \definecolor{YELLOW}{cmyk}{0,0,1,0}
}

\usepackage{babel}
\usepackage{ifpdf}\usepackage{bm}

\makeatother

\begin{document}
\title{Nonlinear and Thermal Effects in the Absorption of Microwaves by Random
Magnets}
\author{Dmitry A. Garanin and Eugene M. Chudnovsky}
\affiliation{Physics Department, Herbert H. Lehman College and Graduate School,
The City University of New York, 250 Bedford Park Boulevard West,
Bronx, New York 10468-1589, USA }
\date{\today}
\begin{abstract}
We study the temperature dependence of the absorption of microwaves
by random-anisotropy magnets. It is governed by strong metastability
due to the broad distribution of energy barriers separating different
spin configurations. At a low microwave power, when the heating is
negligible, the spin dynamics is close to linear. It corresponds to
the precession of ferromagnetically ordered regions that are in resonance
with the microwave field. Previously we have shown (\href{http://doi.org/10.1103/PhysRevB.103.214414}{http://doi.org/10.1103/PhysRevB.103.214414})
that in this regime a dielectric substance packed with random magnets
would be a strong microwave absorber in a broad frequency range. Here
we demonstrate that on increasing the power, heating and over barrier
spin transitions come into play, resulting in the nonlinear behavior.
At elevated temperatures the absorption of microwave power decreases
dramatically, making the dielectric substance with random magnets
transparent for the microwaves. 
\end{abstract}
\maketitle

\section{Introduction}

Random-anisotropy (RA) magnets find numerous technological applications.
Their static properties have been intensively studied during the last
four decades, see, e.g, Refs.\ \onlinecite{RA-book,CT-book,PCG-2015}
and references therein. They are materials that have ferromagnetic
exchange interaction between neighboring spins but random directions
of local magnetic anisotropy, either due to their amorphous structure
or due to sintering from randomly oriented nanocrystals. Depending
on the ratio of the ferromagnetic exchange, $J$, and local magnetic
anisotropy $D_{R}$, they can be hard or soft magnets. At $D_{R}\ll J$
they consist of large ferromagnetically oriented regions, often called
Imry-Ma (IM) domains, and are characterized by low coercivity and
high magnetic susceptibility. In the opposite limit of large anisotropy
compared to the exchange they have high coercivity and large area
of the hysteresis loop. The second limit would be difficult to achieve
in conventional ferromagnets due to the spin-orbit relativistic nature
of the magnetic anisotropy. However, in the cintered RA magnets it
is the effective magnetic anisotropy, $\tilde{D}_{R}$, that enter
the problem. Since $\tilde{D}_{R}$ goes up with the average size
of the nanocrystals (or amorphous structure factor), both limits can
be easily realized in experiment.

The case of strong effective magnetic anisotropy is conceptually simple.
It is equivalent to the array of densely packed, randomly oriented,
single-domain magnetic particles. However, the case of $D_{R}\leq J$
has been subject of a significant controversy. It was initially analyzed
in terms of the IM argument \cite{IM,CSS-1986} that explores the
analogy with the random walk problem: Weak random local pushes from
the RA make the direction of the magnetization created by the strong
exchange interaction wonder around the magnet with the ferromagnetic
correlation length $R_{f}/a\propto(J/D_{R})^{2/(4-d)}$ ($a$ being
the interatomic distance and $d$ being the dimensionality of the
system). The validity of this argument that ignores metastable states,
was later questioned by numerical studies \cite{SL-JAP1987,DB-PRB1990,DC-1991}.
It was found that the RA magnets exhibit metastability and history
dependence \cite{nonergodic,GC-EPJ} regardless of the strength of
the RA, although the IM argument roughly holds for the average size
of ferromagnetically correlated region if one begins with a fully
disordered state. More recently, using random-field model, it was
demonstrated that the presence of topological defects determined by
the relation between the number of spin components and dimensionality
of space, is responsible for the properties of random magnets \cite{GCP-PRB2013,PGC-PRL,CG-PRL}.
Nevertheless, questions about the exact ground state, spin-spin correlations,
topological defects, etc. in the RA model remain largely unanswered
after a 40-year effort.

In the absence of rigorous theory describing static properties of
the RA ferrromagnet, the studies of the dynamics present a significant
challenge. Collective modes and their localization have been observed
in amorphous ferromagnets with random local magnetic anisotropy \cite{Suran-RA,Suran-EPL,Suran-localization,Suran-PRB1997,Suran-JAP1998},
inhomogeneous thin magnetic films \cite{McMichael-PRL2003}, submicron
magnetic heterostructures \cite{Loubens-PRL2007}, and in films where
inhomogeneous magnetic field was generated by a tip of a force microscope
\cite{Du-PRB2014}. Complex nature of these excitations that involves
longitudinal, transversal, and mixed modes has been reported. Following
these experiments, analytical theory of the uniform spin resonance
in a thin film of the RA ferromagnet in a nearly saturating magnetic
field has been developed \cite{Saslow2018}. The dependence of the
frequency of the longitudinal resonance on the magnetic field and
its angle with the film have been obtained and compared with experimental
findings.

Practical applications of random magnets as absorbers of microwave
power typically involve zero static magnetic field. This problem is
more challenging than the the problem with the saturating field because
the underlying magnetic state is strongly disordered, resembling spin
glasses. Zero-field resonances were observed in spin glasses in the
past \cite{Monod,Prejean,Alloul1980,Schultz,Gullikson}. They were
attributed \cite{Fert,Levy,Henley1982} to the RA arising from Dzyaloshinskii-Moriya
interaction and analyzed within hydrodynamic theory \cite{HS-1977,Saslow1982}.
Due to the lack of progress on spin glasses and random magnets, theoretical
effort aimed at understanding their dynamics was largely abandoned
in 1990s. Recently the authors returned to this problem using the
power of modern computers within the framework of the RA magnet in
zero magnetic field \cite{GC-PRB2021}. Images of spin oscillations
induced by the ac field have been obtained. They confirmed earlier
conjectures that collective modes in the RA ferromagnet were localized
within correlated volumes that are in resonance with the ac field.
It was shown that broad distribution of resonances makes such magnets
excellent broadband absorbers of the microwave power that can compete
with nanocomposites commonly used for that purpose \cite{nanocomposites,carbon}.

In Ref.\ \onlinecite{GC-PRB2021} the power of the microwave radiation
and the duration of the radiation pulse was assumed to be very weak
to cause any heating or significant nonlinearity in the response of
the system consisting, e.g., of a dielectric layer packed with RA
magnets. Meanwhile, in certain applications such a system would be
subjected to a strong microwave beam that might cause nonlinear effects
and heating. Here we address the question of how they would influence
absorption. We show that in a RA magnet, the absorbed energy is quickly
redistrubuted over all degrees of freedom, so that the magnetic system
is in equilibrium and the nonlinearity and saturation of the absorption
can be effectively described in terms of heating. Then the task reduces
to computing the temperature dependence of the power absorption. This
allows to describe the whole process in terms of a single differential
equation for the time dependence of the spin temperature. We use Monte
Carlo method to prepare initial states at different temperatures,
then run conservative dynamics and calculate the absorbed power based
upon fluctuation-dissipation theorem. (FDT). Frequency dependence
of the power shows evolution from a broad absorption peak at low temperatures
to the low-absorption plateau on increasing temperature. 

Practically, there is energy transfer from the spins to the atomic
lattice and further to the dielectric matrix containing the RA magnets
and to the substrate that it is deposited on. The transfer of energy
from spins to other degrees of freedom they are interacting with,
such as, e.g., phonons, is fast and effectively increases the heat
capacity of the system, reducing heating to some extent. The heat
flow from the magnets to the dielectric matrix and then to the substrate,
that can significantly reduce the heating of the spin system by microwaves,
is much slower. Here we consider microwave pulses of duration that
is sufficient to heat the RA magnet but short compared to the typical
times of the heat flow out of the RA magnet due to thermal conductivity.
We show that sufficiently strong and long microwave pulses are heating
RA magnets and make them transparent for the microwaves.

The paper is organized as follows. The RA model and relevant theoretical
concepts are introduced in Section \ref{Sec_Theory}. Sec. \ref{Sec_Numerical-procedures}
describes the two numerical experiments performed in this work: (i)
pumping the system by an ac field and (ii) running conservative dynamics
at different temperatures and computation the absorbed power via FDT.
Sec. \ref{Sec_Numerical-results} reports the results of these two
experiments. Our results and possible applications are discussed in
Section \ref{Sec_Discussion}.

\section{Theory}

\label{Sec_Theory}

Following Ref.\ \onlinecite{GC-PRB2021} we consider a model of
three-component classical spin vectors ${\bf s}_{i}$ on the lattice
described by the Hamiltonian



\begin{equation}
\mathcal{H}=-\frac{1}{2}\sum_{i,j}J_{ij}\mathbf{s}_{i}\cdot{\bf s}_{j}-\frac{D_{R}}{2}\sum_{i}({\bf n}_{i}\cdot{\bf s}_{i})^{2}-\mathbf{h}(t)\cdot\sum_{i}{\bf s}_{i}.\label{Hamiltonian}
\end{equation}
Here the first term is the exchange interaction between nearest nearest
neighbors with the coupling constant $J>0$, $D_{R}$ is the strength
of the easy axis RA in energy units, ${\bf n}_{i}$ is a three-component
unit vector having random direction at each lattice site, and $\mathbf{h}(t)={\bf h}_{0}\sin(\omega t)$
is the ac magnetic field in energy units. We study situations when
the wavelength of the electromagnetic radiation is large compared
to the size of the RA magnet, so that the time-dependent field acting
on the spins is uniform across the system. In all practical cases,
$h_{0}$ is small in comparison to the other terms of the Hamiltonian,
so that the only type of nonlinearity can be saturation of resonances.

We assume \cite{CSS-1986,RA-book,PGC-PRL} that the terms taken into
account are typically large compared to the dipole-dipole interaction
(DDI) between the spins that we neglect. The study of the RA magnets
requires systems of size greater than the ferromagnetic correlation
length. Adding DDI to such problem considerably slows down the numerical
procedure without changing the results in any significant way because
the IM domains generated by the RA are much smaller than typical domains
generated by the DDI \cite{CT-book}.

In Ref.\ \onlinecite{GC-PRB2021} it was shown that the microwave
absorption is qualitatively similar in RA systems of different dimensions.
With that in mind and to reduce computation time that becomes prohibitively
long for a sufficiently large 3D system, we do most of the numerical
work for the 2D model on a square lattice.

We consider conservative dynamics of the magnetic system governed
by the dissipationless Landau-Lifshitz (LL) equation 
\begin{equation}
\hbar\dot{\mathbf{s}}_{i}=\mathbf{s}_{i}\times{\bf h}_{{\rm eff},i},\quad{\bf h}_{{\rm eff},i}\equiv-\frac{\partial\mathcal{H}}{\partial\mathbf{s}_{i}}=\mathbf{h}(t)+\mathbf{H}_{\mathrm{eff},i},\label{Larmor}
\end{equation}
where $\mathbf{H}_{\mathrm{eff},i}=\sum_{j}J_{ij}\mathbf{s}_{j}+D_{R}({\bf n}_{i}\cdot{\bf s}_{i}){\bf n}_{i}$
is the effective field due to the exchange and RA. It was shown in
Ref.\ \onlinecite{GC-PRB2021} that including a small dissipation
due to the interaction of the magnetic system with other degrees of
freedom does not change the results for the absorbed power since the
RA magnet has a continuous spectrum of resonances and the system has
its own internal damping due to the many-particle nature of the Hamiltonian,
Eqs.\ (\ref{Hamiltonian}). For the conservative system, the time
derivative of $\mathcal{H}$ is equal to the absorbed power:
\begin{equation}
\dot{\mathcal{H}}(t)=P_{\mathrm{abs}}=-\dot{\mathbf{h}}(t)\cdot\sum_{i}\mathbf{s}_{i}(t).\label{Absorption}
\end{equation}

An important component of the theory of classical spin systems is
the dynamical spin temperature \cite{Nurdin}. For the RA magnet the
general expression given by Ref.\ \onlinecite{Nurdin} yields
\begin{equation}
T_{S}=\frac{\sum_{i}\left(\mathbf{s}_{i}\times\mathbf{H}_{\mathrm{eff},i}\right)^{2}}{2\sum_{i,j}J_{ij}\mathbf{s}_{i}\cdot{\bf s}_{j}+D_{R}\sum_{i}\left[3\left(\mathbf{n}_{i}\cdot\mathbf{s}_{i}\right)^{2}-1\right]}.\label{TS}
\end{equation}
At $T=0$, spins are aligned with their effective fields, and the
numerator of this formula vanishes. At $T=\infty$, spins are completely
disordered, and both terms in the denominator vanish. In a thermal
state of a large system with the temperature $T$, created by Monte
Carlo, $T_{S}\cong T$ up to small fluctuations.

At low temperatures and weak RA, $D_{R}\ll J$, spins are strongly
correlated within large regions of the characteristic size $R_{f}$
(ferromagnetic correlation radius) defined by \cite{IM,CSS-1986}
\begin{equation}
\frac{R_{f}}{a}\sim\left(\frac{J}{D_{R}}\right)^{2/(4-d)},\label{Rf_IM}
\end{equation}
where $a$ is the lattice spacing. In 3D, $R_{f}$ becomes especially
large that makes computations for systems with linear size $L\gg R_{f}$
problematic. Assuming that there is no long-range order (that is the
case when the magnetic state is obtained by energy minimization from
a random state), one can estimate $R_{f}$ using the value of the
system's average spin
\begin{equation}
\mathbf{m}=\frac{1}{N}\sum_{i}\mathbf{s}_{i}
\end{equation}
that is nonzero in finite-size systems. One has
\begin{equation}
m^{2}=\frac{1}{N^{2}}\sum_{i,j}\mathbf{s}_{i}\cdot{\bf s}_{j}=\frac{1}{N}\sum_{j}\langle{\bf s}_{i}\cdot{\bf s}_{i+j}\rangle\Rightarrow\frac{1}{N}\intop_{0}^{\infty}\frac{d^{d}r}{a^{d}}G(r),
\end{equation}
where $G(r)$ is the spatial correlation function and $d$ is the
dimensionality of the space. As the RA magnet has lots of metastable
local energy minima, $G(r)$ depends on the initial conditions and
on the details of the energy minimization routine. In 2D for $G(r)=\exp\left[-\left(r/R_{f}\right)^{p}\right]$
one obtains
\begin{equation}
m^{2}=K_{p}\frac{\pi R_{f}^{2}}{Na^{2}}\quad\Longrightarrow\quad\frac{R_{f}}{a}=m\sqrt{\frac{N}{\pi K_{p}}},
\end{equation}
where $K_{1}=2$ and $K_{2}=1$. Having estimated $R_{f}$, one can
find the number of IM domains in the system. In 2D with linear sizes
$L_{x}$ and $L_{y}$ one has $N_{IM}=L_{x}L_{y}/\left(\pi R_{f}^{2}\right)$.
In particular, for a system with $N=300\times340=102000$ spins and
$D_{R}/J=0.3$, energy minimization at $T=0$ starting from a random
spin state yields $m\approx0.2$1, and with $p=2$ one obtains $R_{f}/a\approx37.8$
and $N_{IM}\approx23$. For $D_{R}/J=1$, one obtains $m\approx0.074$
and $R_{f}/a\approx13.3$ that yields $N_{IM}\approx183$. For the
ratio of the $R_{f}$ values one obtains $R_{f}^{(D_{R}=0.3)}/R_{f}^{(D_{R}=1)}\approx2.84$
that is close to the value 3.33 given by Eq. (\ref{Rf_IM}).

The absorbed power in the linear regime at finite temperatures can
be computed using the fluctuation-dissipation theorem (FDT) relating
the imaginary part of the linear susceptibility per spin, defined
as
\begin{equation}
m_{x}(t)=h_{0}\left[\chi'_{x}\left(\omega\right)\sin\left(\omega t\right)-\chi_{x}''\left(\omega\right)\cos\left(\omega t\right)\right],
\end{equation}
($\mathbf{h}_{0}=h_{0}\mathbf{e}_{x}$) with the Fourier transfom
of the autocorrelation function of the average spin:
\begin{equation}
\chi''{}_{x}(\omega)=\frac{\omega N}{k_{B}T}\mathrm{Re}\int_{0}^{\infty}dt\,e^{i\omega t}A_{x}(t),\label{chi_FDT}
\end{equation}
where
\begin{equation}
A_{x}(t)\equiv\langle[m_{x}(t)-\langle m_{x}\rangle][m_{x}(0)-\langle m_{x}\rangle]\rangle.\label{ACF_def}
\end{equation}
The absorbed power per spin is related to $\chi''_{x}(\omega)$ as
\begin{equation}
P_{\mathrm{abs}}(\omega)=\frac{1}{2}\omega\chi''_{x}(\omega)h_{0}^{2}.
\end{equation}
 Because of the overall isotropy, one can symmetrize over directions
and use
\begin{equation}
\frac{P_{\mathrm{abs}}(\omega)}{h_{0}^{2}}=\frac{\omega^{2}N}{2k_{B}T}\mathrm{Re}\int_{0}^{\infty}dt\,e^{i\omega t}A(t),\label{P_abs_FDT}
\end{equation}
where $A(t)\equiv\left\langle \mathbf{m}(t)\cdot\mathbf{m}(0)\right\rangle /3.$
Here in the definition of $A(t)$ the subtraction terms were dropped
as they do not make a contribution at finite frequencies.

\section{Numerical procedures}

\label{Sec_Numerical-procedures}

In this work, all computations were done for a 2D model of $300\times340=102000$
spins on a square lattice for $D_{R}/J=1$. The frequency-dependent
absorbed power for this model at $T=0$ was computed in Ref.\ \onlinecite{GC-PRB2021}
by pumping the system with ac fields of different frequencies. Here
two numerical experiments were done.

The first experiment was pumping of the system prepared at $T=0$
by the energy minimization \cite{GCP-PRB2013} starting from a random
state by the ac field of different amplitudes $h_{0}$ at a fixed
frequency $\omega/J=0.075$ corresponding to the absorption maximum.
The pumping routine was very long to study nonlinearity, resonance
saturation, and system heating. Eq. (\ref{Absorption}) was used to
compute the absorbed power.

The second experiment was extracting, using FDT, of the absorbed power
$P_{\mathrm{abs}}(\omega.T)$ in the linear regime from the conservative
dynamics of the system prepared by Monte Carlo at different temperatures.
Using the direct method of pumping at nonzero temperature is problematic
as fluctuations of $\mathbf{m}$ in a finite-size system easily dominate
the response to a weak ac field $\mathbf{h}(t)$. Obtaining good results
requires a very long computation to make these fluctuations average
out or large values of the ac amplitude $h_{0}$ that makes the response
nonlinear. Using FDT, one obtains exactly the linear response, and
one computation of the conservative time evolution yields the results
at all frequencies. As averaging of fluctuations is a part of this
procedure as well, a long dynamical evolution is needed. 

Computing a long dynamical evolution of conservative systems requires
an ordinary differential-equation (ODE) solver that conserves energy.
Currently, many researchers prefer simplectic solvers that conserve
energy exactly. However, the precision of popular simplectic solvers
is low, step error $\delta t^{3}$, while higher order symplectic
methods are cumbersome. In addition, these methods do not work for
systems with single-site anisotropy. On the other hand, mainstream
general-purpose solvers such as Runge Kutta 4 (step error $\delta t^{5}$)
and Runge Kutta 5 (step error $\delta t^{6}$) do not conserve the
energy explicitly. Although the energy drift is small, it accumulates
over large computation times. 

\begin{figure}[h]
\centering{}\includegraphics[width=9cm]{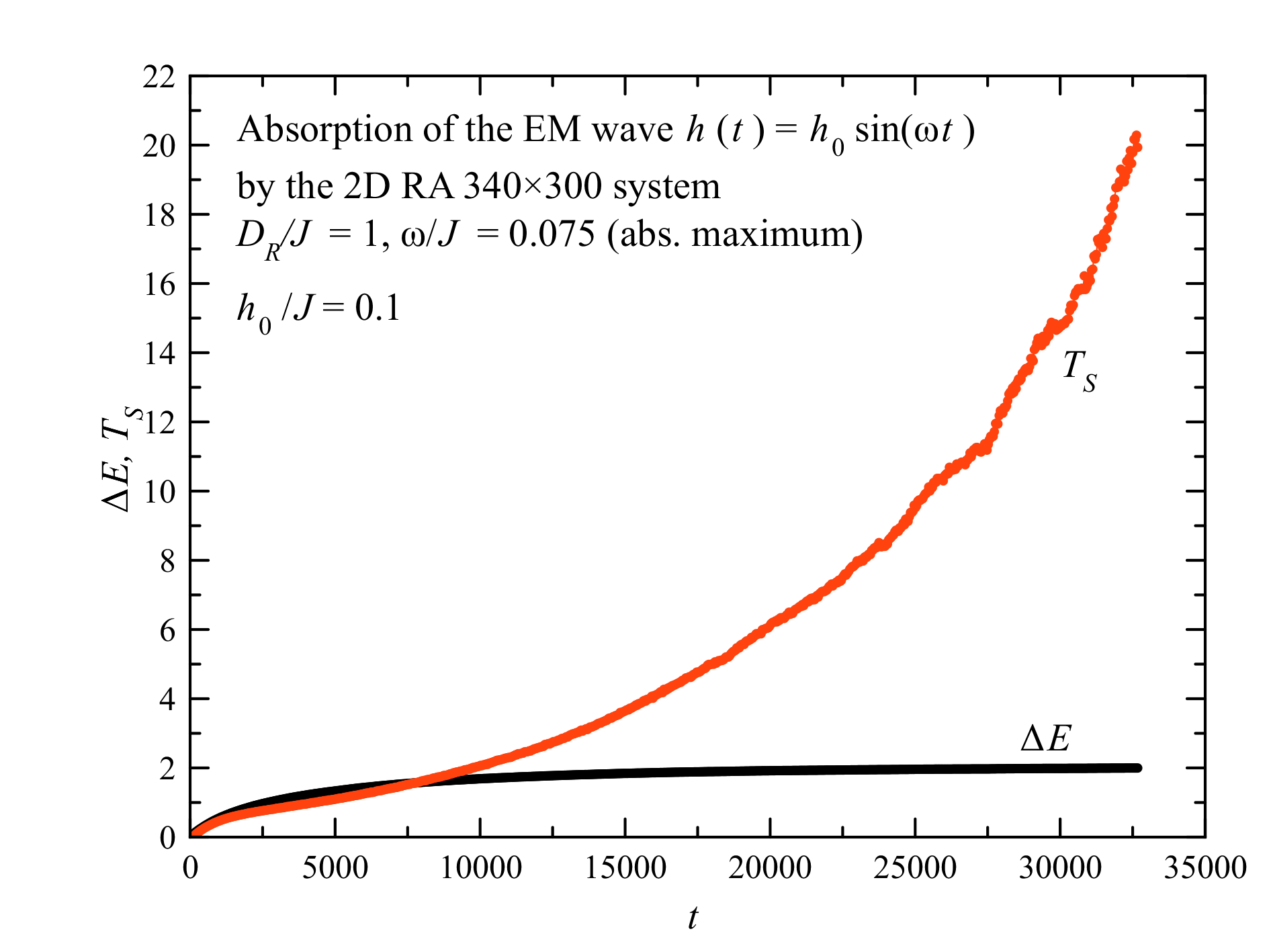} \caption{Time dependence of the absorbed microwave energy $\Delta E$ and of
the spin temperature $T_{s}$ in the RA ferromagnet.}
\label{Fig_DeltaE-t}
\end{figure}

\begin{figure}[h]
\centering{}\includegraphics[width=9cm]{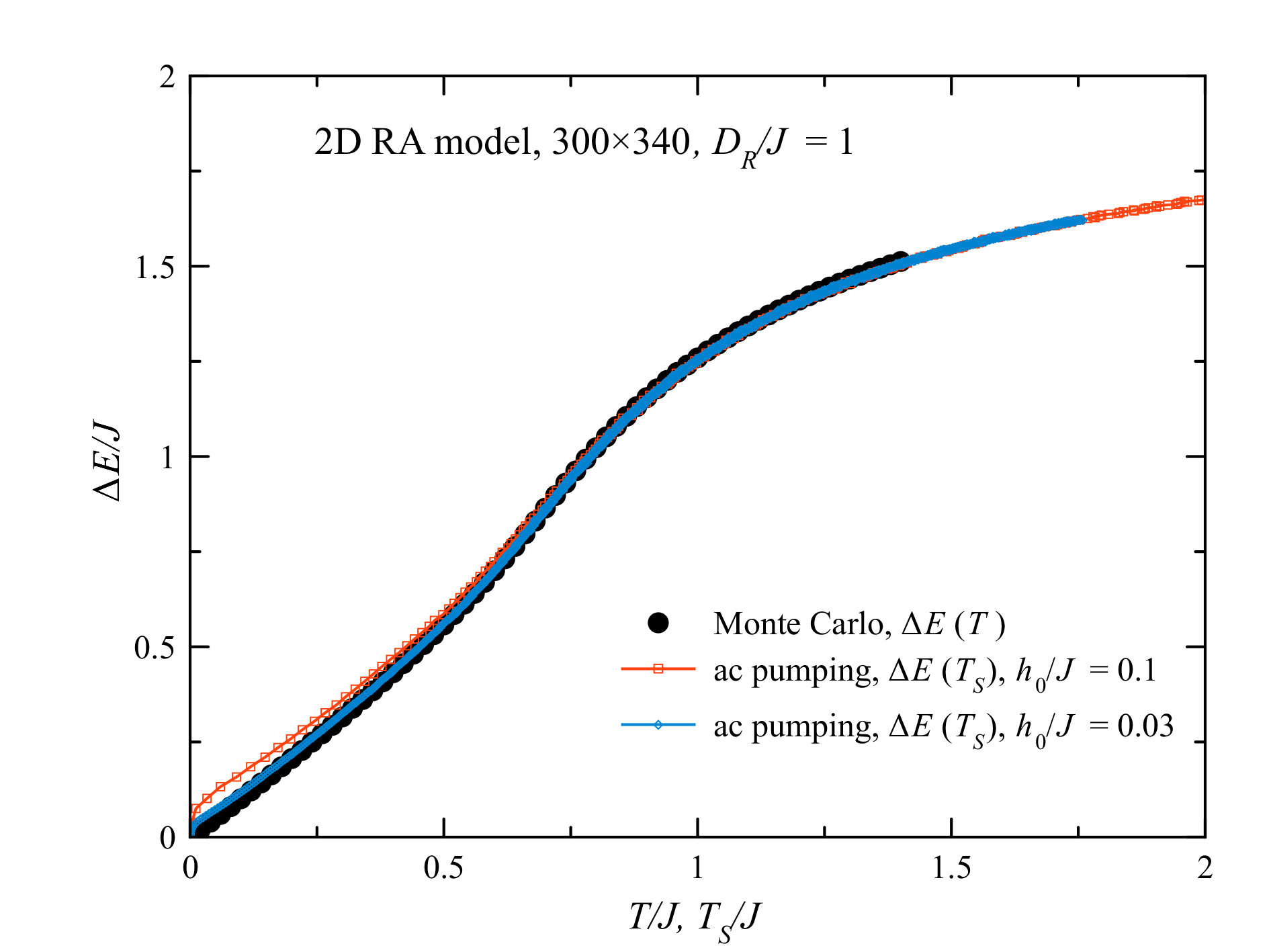}\caption{Absorbed energy vs Monte Carlo temperature $T$ and the spin temperature
$T_{s}$.}
\label{Fig_E-Ts}
\end{figure}

This inconvenience can be cured by the recently proposed procedure
of energy correction \cite{Garanin} that can be performed from time
to time to bring the system's energy back to the expected target value.
This procedure consists in a (small) rotation of each spin towards
or away from the respective effective field

\begin{equation}
\delta{\bf s}_{i}=\xi{\bf s}_{i}\times({\bf s}_{i}\times{\bf H}_{{\rm eff},i}),
\end{equation}
with $\xi$ chosen such that the ensuing energy change $\delta E=\xi\sum_{i}({\bf s}_{i}\times{\bf H}_{{\rm eff},i})$
has a required value. In the case of a free conservative dynamics,
one has $\delta E=E_{0}-E$, where $E_{0}$ is the initial energy
that must be conserved and $E$ is the current energy that differes
from $E_{0}$ because of the accumulation of numerical errors.

In our first experiment with the ac pumping, the situation with the
energy conservation is the following. The absorbed energy $E_{\mathrm{abs}}(t)=\int_{0}^{t}dt'P_{\mathrm{abs}}(t')$
is robust because any portion of the absorbed energy that has been
counted does not change. On the contrary, the energy of the system
$E\equiv\mathcal{H}$ changes because of numerical errors, even in
the absence of pumping. If it changes significantly, then it begins
to affect the absorbed power and the whole computation breaks down.
Thus the energy of the system must be corrected so that the energy
change $\Delta E(t)=E(t)-E_{0}$ is equal to the absorbed energy $E_{\mathrm{abs}}(t)$.
This implies $\delta E=E_{\mathrm{abs}}-\Delta E$ in the energy-correcting
transformation. In the case of pumping, this transformation should
be done at the time moments when $\mathbf{h}(t)=0$, to avoid the
ac field making a contribution to the energy. In this work, it was
done each time a period of the pumping was completed.

In our second experiment with the free conservative dynamics and FDT,
energy correction also was done with fixed time intervals. In this
experiment, however, one could do a Monte Carlo update instead of
correcting energy, to remove the energy drift. In both experiments,
the computation time was not limited by any accuracy factors.

As the ODE solver, the fifths-order Runge-Kutta method with the time
step $\delta t=0.1$ was used. As the computational tool, we employed
Wolfram Mathematica with compilation of heavy-duty routines into C++
code using a C compiler installed on the computer. Most of the computations
were performed on our Dell Precision Workstation having 20 cores.
The computation in the first numerical experiment was not parallelized.
For the second experiment, to better deal with fluctuations, we ran
identical parallel computations on 16 cores available under our Mathematica
licence and averaged the results. In computations, we set $J$ and
all physical constants to one. The time $t$ is given in the units
of $\hbar/J$.

\section{Numerical results}

\label{Sec_Numerical-results}

\begin{figure}[h]
\begin{centering}
\includegraphics[width=9cm]{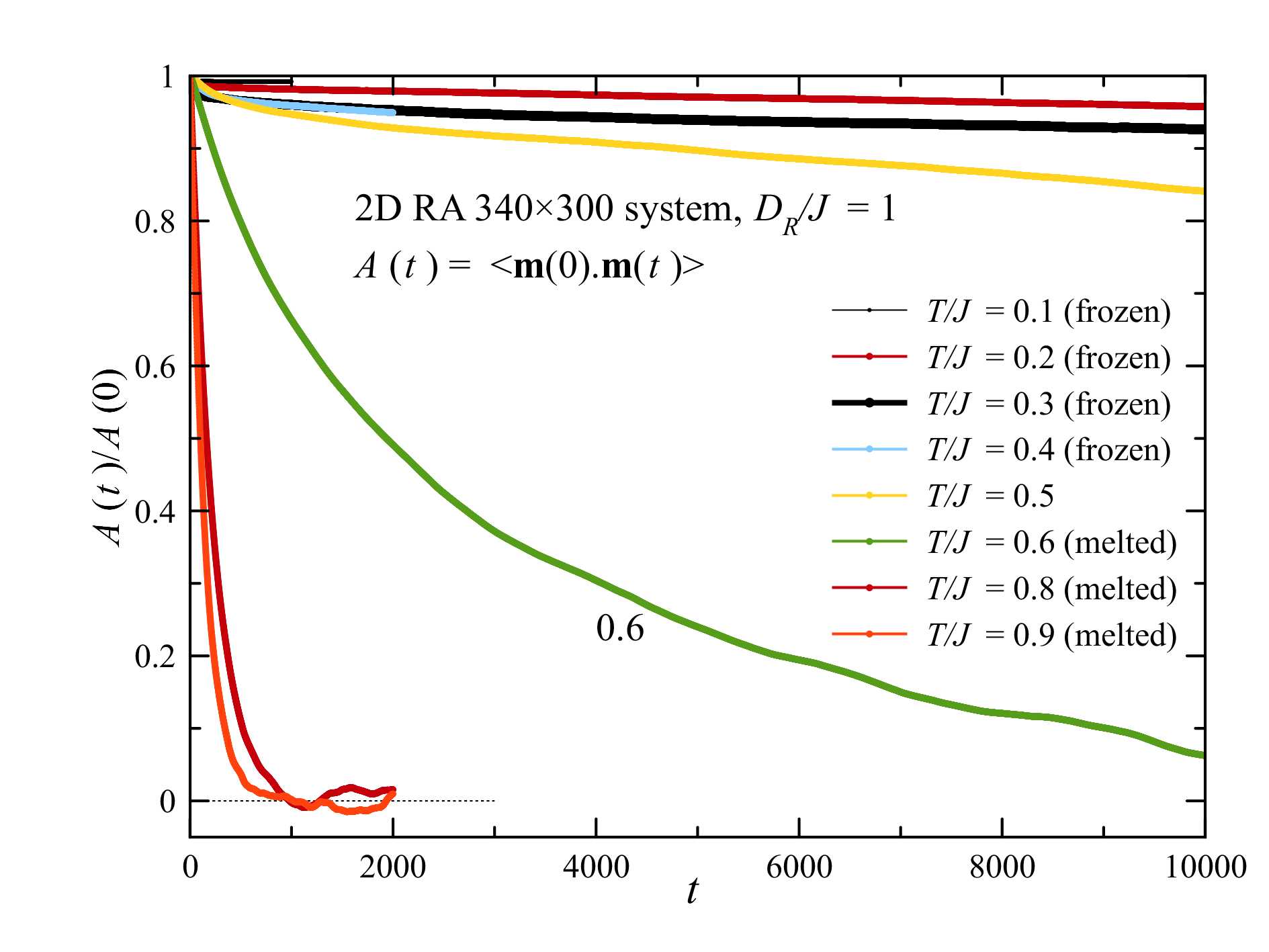}
\par\end{centering}
\begin{centering}
\includegraphics[width=9cm]{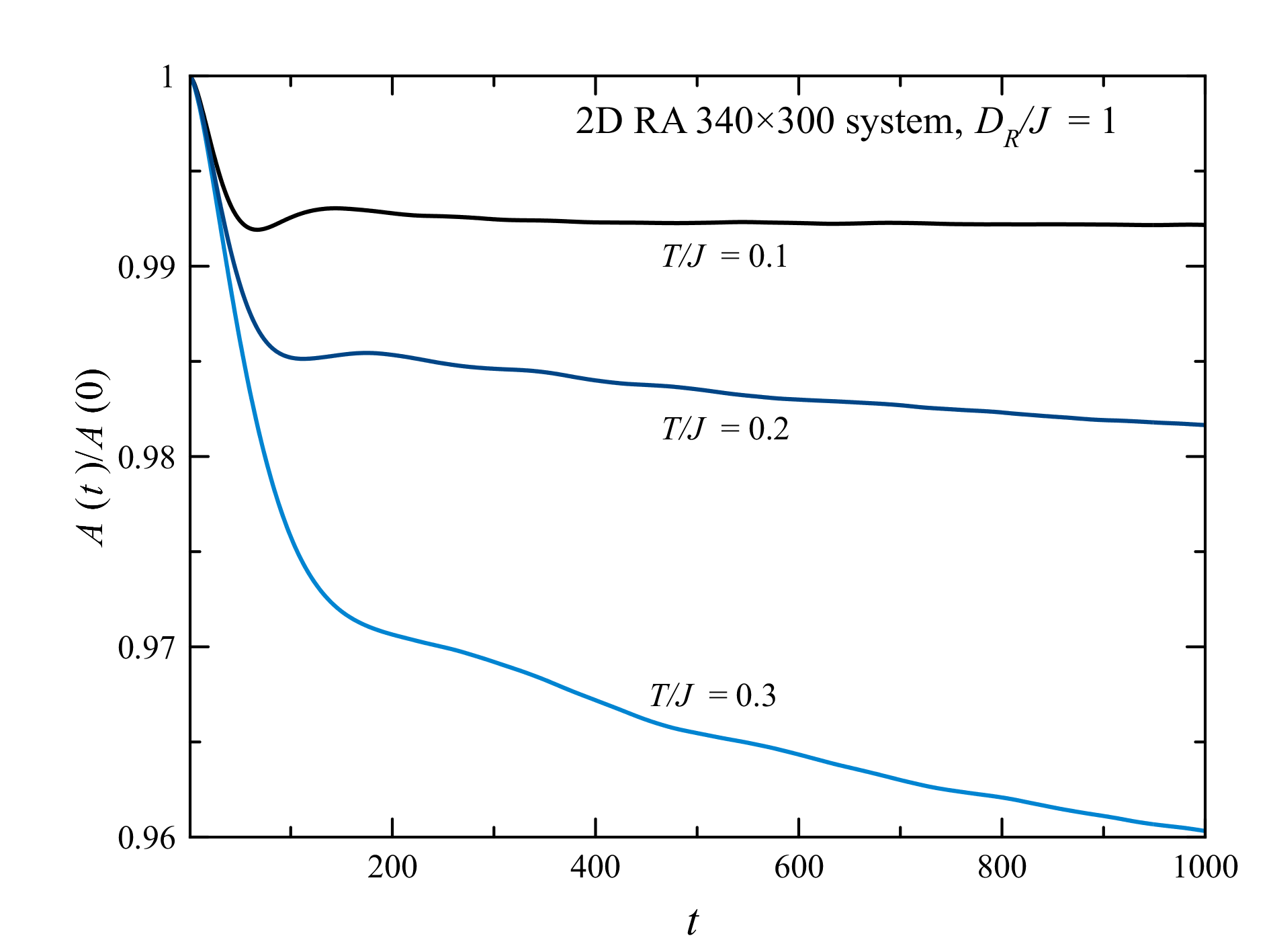}
\par\end{centering}
\centering{}\caption{Autocorrelation function of the total spin at different temperatures
temperatures. Upper panel: wide temperature range. Lower panel: low
temperatures.}
\label{Fig_A_t}
\end{figure}

\subsection{Pumping, nonlinearity, heating}

\label{Sec_pumping}

Pumping an isolated system during a long time increases its energy,
that is, results in heating. For realistically small ac amplitudes,
$h_{0}/J\lll1$, the computation times required to reach a significant
heating are prohibitively long. A spectacular heating can be reached
for the extremely high ac amplitude $h_{0}/J=0.1$, although it requires
a very long computation time. The results are shown in Fig. \ref{Fig_DeltaE-t}.
At long times, the energy change of the system $\Delta E$ (equal
to the absorbed energy $E_{\mathrm{abs}}$) reaches its maximal value
corresponding to a total disordering of spins. In this state, the
absorbed power is close to zero and the saturation is complete. The
dynamical spin temperature $T_{S}$ defined by Eq. (\ref{TS}) reaches
large values, as it should be. For $h_{0}/J=0.03$ that is also too
large, a very long computation allowed to reach the temperature $T_{S}\approx1.75J$. 

The existence of the dynamic spin temperature $T_{S}$ allows to check
if the state of the system is equilibrium during the energy-absorption
process. In Fig. \ref{Fig_E-Ts} the results above for $\Delta E(t)$
and $T_{S}(t)$ are parametrically replotted and compared with the
dependence $\Delta E(T)$ computed by Monte Carlo. The parametric
plots $\Delta E(T_{S})$ for $h_{0}/J=0.1$ and 0.03 perfectly coincide
everywhere except the smallest $T_{S}$ where the $h_{0}/J=0.1$ curve
bulges. For the $h_{0}/J=0.03$ the bulging also exists but is rather
small. Both of these curves are in a good accord with the Monte Carlo
curve $\Delta E(T)$. This result strongly suggests that as the RA
magnet is absorbing the energy, the latter goes to all modes and the
equipartition of the energy is reached. It is remarkable that the
system is in equilibrium even for the huge ac amplitude $h_{0}/J=0.1$,
except for the short times. For the realistically weak ac amplitudes,
the equilibrium in the magnetic system should be complete.

The magnetic system being in equilibrium and having a particular spin
temperature $T_{S}$ during the absorption of the microwave energy
allows one to set up a single differential equation for the temperature
that describes the whole process. Expressing the time derivative of
the system's energy in Eq. (\ref{Absorption}) via the heat capacity
of the magnetic system $C$ and the time derivative of the spin temperature,
one obtains
\begin{equation}
\dot{\mathcal{H}}(t)=C\dot{T}=P_{\mathrm{abs}}(\omega,T)
\end{equation}
that results in a closed equation for $T(t)$
\begin{equation}
\dot{T}=\frac{P_{\mathrm{abs}}(\omega,T)}{C(T)}.\label{TS_eq}
\end{equation}
Thus, it is sufficient to compute the heat capacity of the spin system
$C(T)$ by Monte Carlo and compute the absorbed power $P_{\mathrm{abs}}(\omega.T)$
at different temperatures to be able to solve the problem of heating
and the resonance saturation numerically in no time! As the realistic
ac field amplitudes are rather small, one can find $P_{\mathrm{abs}}(\omega.T)$
in the linear regime with the help of FDT. After the numerical solution
of Eq. (\ref{TS_eq}) is obtained, one can compute the absorbed energy
by integration:
\begin{equation}
E_{\mathrm{abs}}(t)=\int_{0}^{t}dt'P_{\mathrm{abs}}\left[\omega,T(t')\right].
\end{equation}

As the spin-lattice relaxation is rather fast, one can add the heat
capacity of the lattice to that of the magnetic system. This will
alleviate the negative action of heating and increase absorption.
Also, the energy can flow from the magnetic particles to the dielectric
matrix by heat conduction. However, this process is slow and it cannot
transfer a significant energy during the time of the microwave pulse.

\subsection{Microwave power absorption by FDT}

\label{Sec_FDT}

In the second numerical experiment, we ran conservative dynamical
evolution of the states created by Monte Carlo at the temperatures
$T/J=0.1\div1.0$ in step of 0.1. The computation was performed in
parallel on 16 processor cores until $t=t_{\max}=50000$ in most cases.
From the computed dependences $\mathbf{m}(t)$ the autocorrelation
function $A(t)$ entering Eq. (\ref{P_abs_FDT}) was computed and
the results were averaged over the cores. Computations on our Dell
Precision Workstation took five days for each temperature. The normalized
autocorrelation functions $A(t)/A(0)$ at different temperatures are
shown in Fig. \ref{Fig_A_t}. These dependences have different forms
at low and elevated temperatures that can be interpreted in terms
of glassy physics. Below $T/J\approx0.6$, spins are frozen and $A(t)$
is decreasing very slowly, apparently due to bunches of spins (IM
domains) crossing anisotropy barriers due to thermal agitation. 

There is a fundamental unsolved question whether the RA system can
be described in terms of IM domains whose magnetic moments become
blocked at low temperatures due to barriers or it is a true spin-glass
state. \cite{Shand-JAP2005} Since our focus is on the absorption
of microwaves we do not attempt to answer this question here. In Eq.
(\ref{P_abs_FDT}) for the absorbed power, the low-frequency part
of $A(\omega)$ is suppressed by the factor $\omega^{2}$, so that
the long-time physics is irrelevant for the absorption. 

The contribution to the absorbed power comes from the short-time part
of $A(t)$ that is shown in the lower panel of Fig. \ref{Fig_A_t}.
There is an initial steep descent of $A(t)$ ending in a quasi-plateau
at low temperatures. This steep descent can be interpered as caused
by dephasing of precession of different IM domains in their potential
wells. This precession with a quasi-continuous spectrum of frequencies
is what ensures the absorption of the microwave power in a broad frequency
range. As each IM domain remains precessing in its own potential well,
$A(t)$ cannot change by a large amount. The latter requires flipping
of IM domains over the barriers that happens at higher temperatures. 

\begin{figure}[h]
\centering{}\includegraphics[width=9cm]{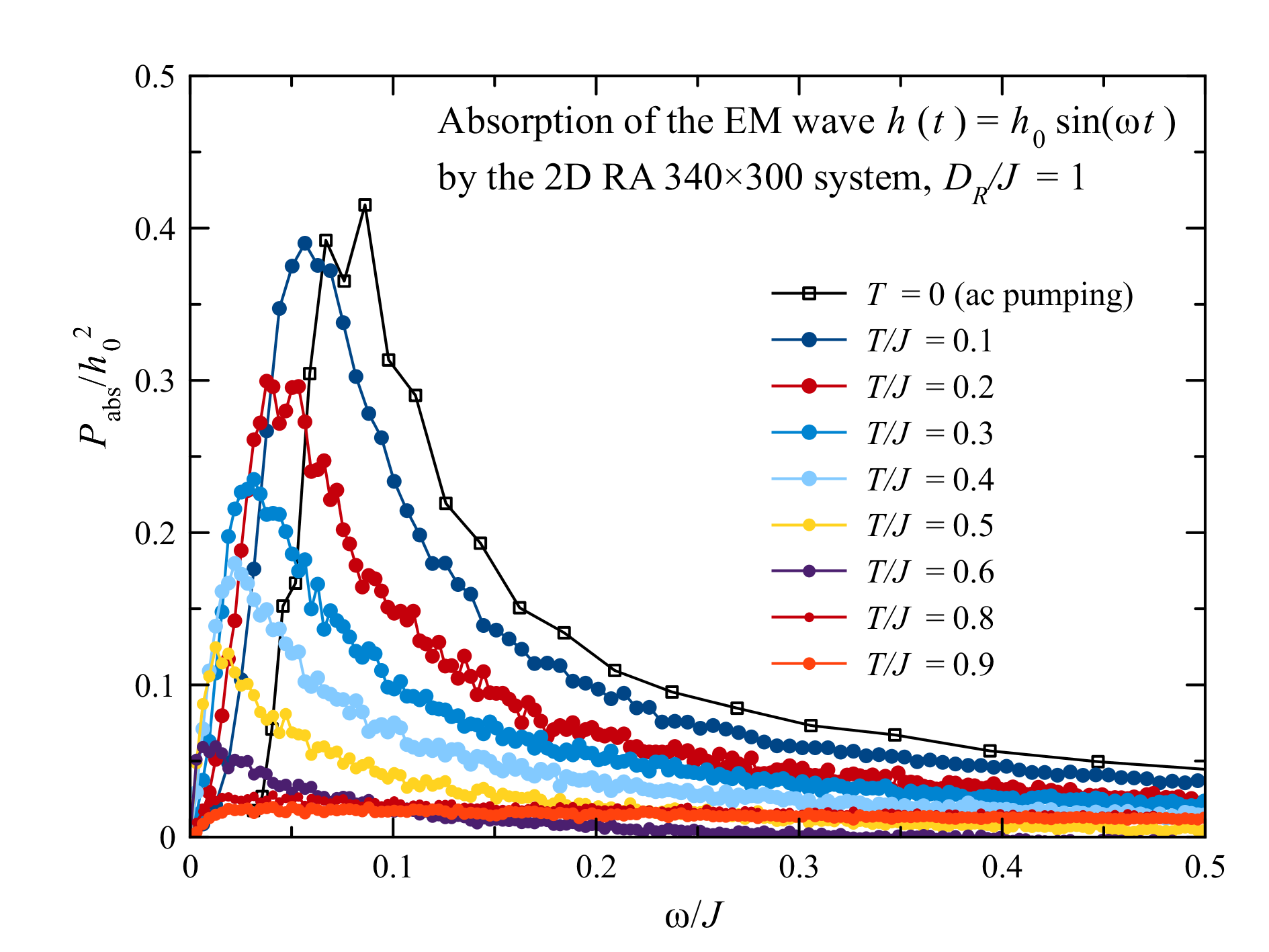}\caption{Frequency dependence of the absorbed microwave power at different
temperatures. The ac pumping result at $T=0$ of Ref.\ \onlinecite{GC-PRB2021}
is shown for comparison. }
\label{Fig_Power_vs_omega_different_T} 
\end{figure}

\begin{figure}
\begin{centering}
\includegraphics[width=9cm]{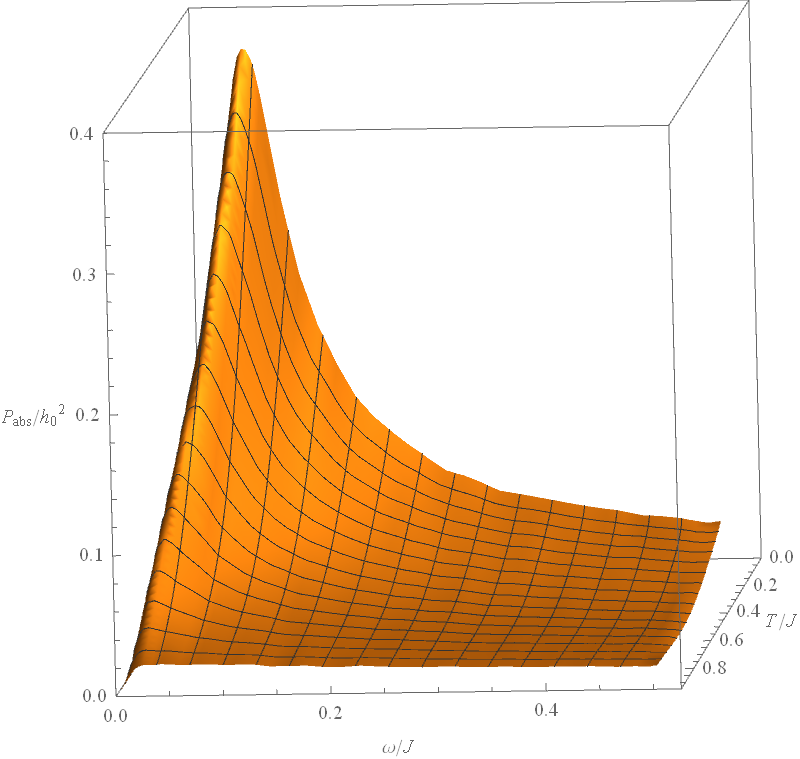}
\par\end{centering}
\caption{Dependence of the absorbed microwave power on the temperature and
frequency.}

\label{Fig_power_vs_omega_and_T}
\end{figure}
\begin{figure}

\begin{centering}
\includegraphics[width=9cm]{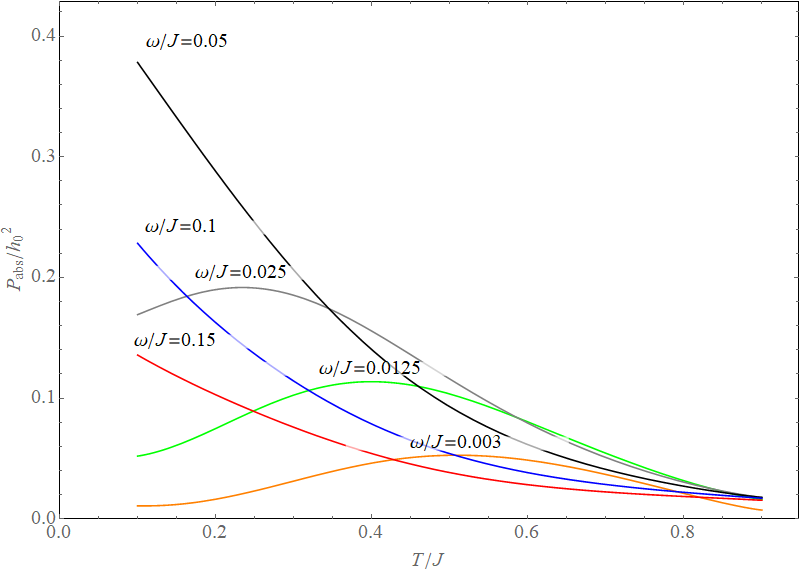}\caption{Temperature dependence of the absorbed microwave power at different
frequencies. }
\label{Fig_power_vs_T_different_omega}
\par\end{centering}
\end{figure}

Making the Fourier transform of $A(t)$ and using Eq (\ref{P_abs_FDT}),
one obtaines the absorbed power $P(\omega)$ that is shown in Fig.
\ref{Fig_Power_vs_omega_different_T}. One can see that the absorption
curve is getting depressed with increasing the temperature, until
the absorption peak vanishes when the glassy state of the system melts.
This is similar to what one observes in a system of independent resonating
spins (or magnetic particles): The more thermally excited are the
spins, the less they absorb. Absorption is maximal if the spins are
in their ground states at $T=0$. 

Fig. \ref{Fig_power_vs_omega_and_T} shows the same data presented
in a 3D form as $P_{\mathrm{abs}}(\omega,T)$ with the help of the
approximation using Bezier functions implemented in Wolfram Mathematica.

Having built the approximation for $P_{\mathrm{abs}}(\omega,T)$,
one can plot the curves $P_{\mathrm{abs}}(T)$ for different $\omega$,
as shown in Fig. \ref{Fig_power_vs_T_different_omega}. Since the
maximum of $P_{\mathrm{abs}}(\omega)$ is shifting to the left with
increasing $T$, the curves $P_{\mathrm{abs}}(T)$ have a maximum
for smaller $\omega$. The dependences $P_{\mathrm{abs}}(T)$ can
be used in the equation for the spin temperature, Eq. (\ref{TS_eq}).

\section{Discussion}

\label{Sec_Discussion}

Earlier we have shown \cite{GC-PRB2021} that RA magnets can be strong
broadband absorbers of the microwave power. In this paper we have
studied the temperature dependence of the power absorption. Our results
answer two questions, both related to applications of microwave absorbers.
The first one is a direct question of how the absorption by RA magnets
depends on their temperature when they are heated by an independent
source. It is answered by Fig. \ref{Fig_Power_vs_omega_different_T}
which shows a decrease of the absorption with increasing temperature
in a broad frequency range, making the system basically transparent
for the microwaves at a sufficiently high temperature. 

The second question is a response of the RA magnet to a microwave
pulse of high power. We have shown that during the pulse, the spin
system is in equilibrium and can be described in terms of the spin
temperature. As typical times of spin-phonon transitions are much
shorter than a microsecond, the spin system also equilibrates with
the lattice. On the other hand, the time required for the flow of
heat out of a dielectric layer of thickness $l$, containing densely
packed RA magnets, can be estimated as $t\sim(\rho cl^{2})/k$, where
$\rho$, $c$, and $k$ are the average mass density, the specific
heat, and the thermal conductivity of the layer. For the layer of
a few-millimeter thickness it is in the ballpark of a fraction of
a second or longer, depending on the substrate. Thus, for the microwave
pulses shorter than this time, the heat-conductivity mechanism is
irrelevant.

A sufficiently strong pulse of microwave energy directed at such a
layer, having duration in the range from microseconds to milliseconds,
can greatly diminish the absorption capacity of the layer during the
action of the pulse, making it transparent for the microwaves. During
that time, if the layer is covering a metallic surface, the microwaves
in a broad frequency range would pass it with the minimum absorption
and would be reflected by the metal. This effect can be minimized
by making the layer densely packed with metallic RA magnets electrically
insulated from each other by a very thin dielectric coating. High
thermal conductivity of such a system would greatly increase its cooling
via heat conduction and would make its absorbing capabilities more
resistant to high-power pulses of direct microwave energy. 

\section*{Acknowledgements}

This work has been supported by the Grant No. FA9550-20-1- 0299 funded
by the Air Force Office of Scientific Research.


\begin{thebibliography}{10}
\bibitem{RA-book} E. M. Chudnovsky, Random Anisotropy in Amorphous
Alloys, Chapter 3 in the Book: \textit{Magnetism of Amorphous Metals
and Alloys}, edited by J. A. Fernandez-Baca and W.-Y. Ching, pages
143-174 (World Scientific, Singapore, 1995).

\bibitem{CT-book} E. M. Chudnovsky and J. Tejada, \textit{Lectures
on Magnetism} (Rinton Press, Princeton, New Jersey, 2006).

\bibitem{PCG-2015} T. C. Proctor, E. M. Chudnovsky, and D. A. Garanin,
Scaling of coercivity in a 3d random anisotropy model, Journal of
Magnetism and Magnetic Materials, \textbf{384}, 181-185 (2015).

\bibitem{IM} Y. Imry and S.-k. Ma, Random-field instability of the
ordered state of continuous symmetry, Physical Review Letters \textbf{35},
1399-1401 (1975).

\bibitem{CSS-1986} E. M. Chudnovsky, W. M. Saslow, and R. A. Serota,
Ordering in ferromagnets with random anisotropy, Physical Review B
\textbf{33}, 251-261 (1986).

\bibitem{SL-JAP1987} R. A. Serota and P. A. Lee, Continuous-symmetry
ferromagnets with random anisotropy, Journal of Applied Physics \textbf{61},
3965-3967 (1987).

\bibitem{DB-PRB1990} B. Dieny and B. Barbara, XY model with weak
random anisotropy in a symmetry-breaking magnetic field, Physical
Review B \textbf{41}, 11549-11556 (1990).

\bibitem{DC-1991} R. Dickman and E. M. Chudnovsky, $XY$ chain with
random anisotropy: Magnetization law, susceptibility, and correlation
functions at $T=0$, Physical Review B \textbf{44}, 4397-4405 (1991).

\bibitem{GCP-PRB2013} D. A. Garanin, E. M. Chudnovsky, and T. Proctor,
Random field xy model in three dimensions, Physical Review B \textbf{88},
224418-(21) (2013).

\bibitem{PGC-PRL} T. C. Proctor, D. A. Garanin, and E. M. Chudnovsky,
Random fields, Topology, and Imry-Ma argument, Physical Review Letters
\textbf{112}, 097201-(4) (2014).

\bibitem{CG-PRL} E. M. Chudnovsky and D. A. Garanin, Topological
order generated by a random field in a 2D exchange model, Physical
Review Letters \textbf{121}, 017201-(4) (2018).

\bibitem{Suran-RA} G. Suran, E. Boumaiz, and J. Ben Youssef, Experimental
observation of the longitudinal resonance mode in ferromagnets with
random anisotropy, Journal of Applied Physics 79, 5381 (1996).

\bibitem{Suran-EPL} S. Suran and E. Boumaiz, Observation and characteristics
of the longitudinal resonance mode in ferromagnets with random anisotropy,
Europhysics Letters \textbf{35}, 615-620 (1996).

\bibitem{Suran-localization} S. Suran and E. Boumaiz, Longitudinal
resonance in ferromagnets with random anisotropy: A formal experimental
demonstration, Journal of Applied Physics \textbf{81}, 4060 (1997).

\bibitem{Suran-PRB1997} G. Suran, Z, Frait, and E. Boumaz, Direct
observation of the longitudinal resonance mode in ferromagnets with
random anisotropy, Physical Review B \textbf{55}, 11076-11079 (1997).

\bibitem{Suran-JAP1998} S. Suran and E. Boumaiz, Longitudinal-transverse
resonance and localization related to the random anisotropy in a-CoTbZr
films, Journal of Applied Physics \textbf{83}, 6679 (1998).

\bibitem{McMichael-PRL2003} R. D. McMichael, D. J. Twisselmann, and
A. Kunz, Localized ferromagnetic resonance in inhomogeneous thin film,
Physical Review Letters \textbf{90}, 227601-(4) (2003).

\bibitem{Loubens-PRL2007} G. de Loubens, V. V. Naletov, O. Klein,
J. Ben Youssef, F. Boust, and N. Vukadinovic, Magnetic resonance studies
of the fundamental spin-wave modes in individual submicron Cu/NiFe/Cu
perpendicularly magnetized disks, Physical Review Letters \textbf{98},
127601-(4) (2007).

\bibitem{Du-PRB2014} C. Du, R. Adur, H. Wang, S. A. Manuilov, F.
Yang, D. V. Pelekhov, and P. C. Hammel, Experimental and numerical
understanding of localized spin wave mode behavior in broadly tunable
spatially complex magnetic configurations, Physical Review B \textbf{90},
214428-(10) (2014).

\bibitem{Saslow2018} W. M. Saslow and C. Sun, Longitudinal resonance
for thin film ferromagnets with random anisotropy, Physical Review
B \textbf{98}, 214415-(6) (2018).

\bibitem{Monod} P. Monod and Y. Berthier, Zero field electron spin
resonance of Mn in the spin glass state, Journal of Magnetism and
Magnetic Materials \textbf{15-18},149-150 (1980).

\bibitem{Prejean} J. J. Prejean, M. Joliclerc, and P. Monod, Hysteresis
in CuMn: The effect of spin orbit scattering on the anisotropy in
the spin glass state, Journal de Physique (Paris) \textbf{41}, 427-435
(1980).

\bibitem{Alloul1980} H. Alloul and F. Hippert, Macroscopic magnetic
anisotropy in spin glasses: transverse susceptibility and zero field
NMR enhancement, Journal de Physique Lettres \textbf{41}, L201-204
(1980).

\bibitem{Schultz} S. Schultz, E .M. Gulliksen, D. R. Fredkin, and
M.Tovar, Simultaneous ESR and magnetization measurements characterizing
the spin-glass state, Physical Review Letters \textbf{45}, 1508-1512
(1980).

\bibitem{Gullikson} E. M. Gullikson, D. R. Fredkin, and S. Schultz,
Experimental demonstration of the existence and subsequent breakdown
of triad dynamics in the spin-glass CuMn, Physical Review Letters
\textbf{50}, 537-540 (1983).

\bibitem{Fert} A. Fert and P. M. Levy, Role of anisotropic exchange
interactions in determining the properties of spin-glasses, Physical
Review Letters \textbf{44},1538-1541 (1980).

\bibitem{Levy} P. M. Levy and A. Fert, Anisotropy induced by nonmagnetic
impurities in CuMn spin-glass alloys, Physical Review B \textbf{23},
4667 (1981).

\bibitem{Henley1982} C. L. Henley, H. Sompolinsky, and B. I. Halperin,
Spin-resonance frequencies in spin-glasses with random anisotropies,
Physical Review B \textbf{25}, 5849-5855, (1982).

\bibitem{HS-1977} B. I. Halperin and W. M. Saslow, Hydrodynamic theory
of spin waves in spin glasses and other systems with noncollinear
spin orientations, Physical Review B \textbf{16}, 2154-2162 (1977).

\bibitem{Saslow1982} W. M. Saslow, Anisotropy-triad dynamics, Physical
Review Letters \textbf{48}, 505-508 (1982).

\bibitem{GC-PRB2021} D. A. Garanin and E. M. Chudnovsky, Absorption
of microwaves by random-anisotropy magnets, Physical Review B \textbf{103},
214414-(11) (2021).

\bibitem{nonergodic} I. Y. Korenblit and E. F. Shender, Spin glasses
and nonergodicity, Soviet Physics Uspekhi \textbf{32}, 139-162 (1989).

\bibitem{GC-EPJ} D. A. Garanin and E. M. Chudnovsky, Ordered vs.
disordered states of the random-field model in three dimensions, European
Journal of Physics B \textbf{88}, 81-(19) (2015).

\bibitem{nanocomposites} J. V. I. Jaakko et al., Magnetic nanocomposites
at microwave frequencies, in a book: \textit{Trends in Nanophysics},
edited by V. Barsan and A. Aldea, pages 257-285 (Springer, New York,
2010).

\bibitem{carbon} X. Zeng, X. Cheng, R. Yu, and G. D. Stucky, Electromagnetic
microwave absorption theory and recent achievements in microwave absorbers,
Carbon \textbf{168}, 606-623 (2020).

\bibitem{Garanin} D. A. Garanin, Energy balance and energy correction
in dynamics of classical spin systems, Condensed Matter arXiv:2106.14689.

\bibitem{Nurdin} W. B. Nurdin and K.-D. Schotte, Dynamical temperature
for spin systems, Physical Review E \textbf{61}, 3579-3582 (2000).

\bibitem{Shand-JAP2005} P. M. Shand, C. C. Stark, D. Williams, M.
A. Morales, T. M. Pekarek, and D. L. Leslie-Pelecky, Spin glass or
random anisotropy?: The origin of magnetically glassy behavior in
nanostructured GdAl$_{2}$, Journal of Applied Physics \textbf{97},
10J505-(3) (2005).
\end{thebibliography}
\end{document}